\documentclass[aip,reprint]{revtex4-1}

\usepackage{graphicx}
\usepackage{dcolumn}
\usepackage{bm}

\usepackage[utf8]{inputenc}
\usepackage[T1]{fontenc}
\usepackage{mathptmx}
\usepackage{etoolbox}

\usepackage{hyperref}
\usepackage[table]{xcolor} 
\hypersetup{
    colorlinks=true,
    linkcolor=blue,
    filecolor=magenta,      
    urlcolor=blue}

\draft 

\begin{document}


\title{Physical origin of enhanced electrical conduction in aluminum-graphene composites} 

\author{K. Nepal}
\email{kn478619@ohio.edu}
\affiliation{Department of Physics and Astronomy, Nanoscale and Quantum Phenomena Institute (NQPI), Ohio University, Athens, Ohio 45701, USA}%

\author{C. Ugwumadu}
\affiliation{Department of Physics and Astronomy, Nanoscale and Quantum Phenomena Institute (NQPI), Ohio University, Athens, Ohio 45701, USA}%

\author{K.N. Subedi}
\affiliation{Theoretical Division, Los Alamos National Laboratory, Los Alamos, New Mexico 87545, USA}%

\author{K. Kappagantula}%
\affiliation{Pacific Northwest National Laboratory, 
Richland, Washington, 99352, USA}%

\author{D. A. Drabold}%
\email{drabold@ohio.edu}
\affiliation{Department of Physics and Astronomy, Nanoscale and Quantum Phenomena Institute (NQPI), Ohio University, Athens, Ohio 45701, USA}%

\date{\today}

\begin{abstract}
The electronic and transport properties of aluminum-graphene composite materials were investigated using \textit{ab initio} plane wave density functional theory.  The interfacial structure is reported for several configurations. In some cases, the face-centered aluminum (111) surface relaxes in a nearly ideal registry with graphene, resulting in a remarkably continuous interface structure. The Kubo-Greenwood formula and space-projected conductivity were employed to study electronic conduction in aluminum single- and double-layer graphene-aluminum composite models. The electronic density of states at the Fermi level is enhanced by the graphene for certain aluminum-graphene interfaces, thus, improving electronic conductivity. In double-layer graphene composites, conductivity varies non-monotonically with temperature, showing an increase between 300-400 K at short aluminum-graphene distances, unlike the consistent decrease in single-layer composites.
\end{abstract}


\maketitle 

Recent experimental research has shown that composites formed by the inclusion of single layer or multiple layers of graphene into aluminum (Al) and copper (Cu) improve the electronic conduction properties of bulk metal.  The interfacial structure of the metal-graphene composites is generally believed to form a high-energy configuration under suitable compression for specific experimental designs, such as hot-extrusion \cite{Aditya_MSE,KAPPAGANTULA2022162477} or friction-extrusion \cite{Gwalani_MD} methods. This discovery holds promise for long-distance power transmission, and other applications \cite{SHARMA20214133, article2, Sauvage_2015}.

\begin{figure*}[t!]
	\includegraphics[width=\textwidth]{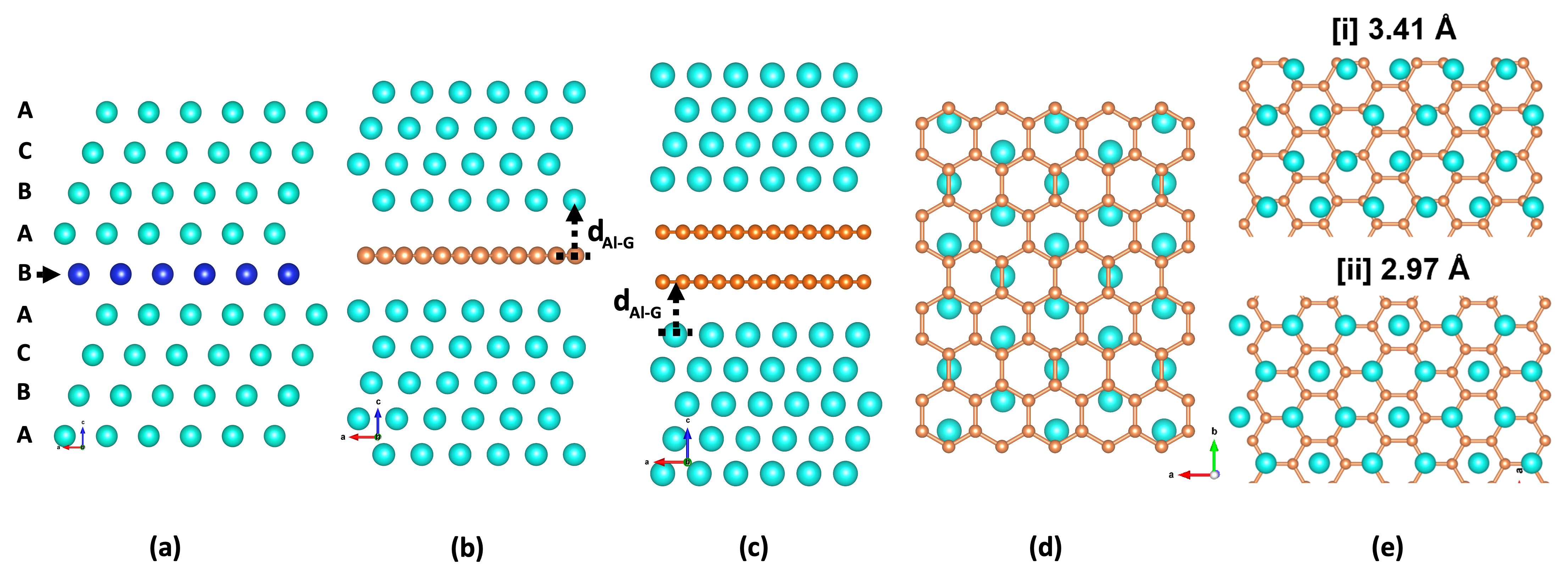}
    \centering
	\caption{(a) ABCABC… planar stacking in the Al (111) face-centered structure with a fault layer, shown by blue atoms. Representative structure of interface models with (b) single-layer and (c) double-layer graphene, where d$_{Al-G}$ is the distance between the aluminum surface and the graphene layer. (d) Top view of the arrangement of carbon atoms in the graphene layer with Al (111) for a weakly interacting Al-graphene system with the interfacial distance of 3.48 \AA~(before relaxation). (e) Interface structure in composite models after the relaxation of DL composites with [i] $d_{Al-G} \ = $ 3.41 and [ii] 2.97 \AA. For $d_{Al-G} \ = $ 2.97 \AA, the optimized interface structure forms a strain-free registry between Al and C. All cyan (brown) spheres represent Al (C) atoms.}
    
	\label{fig:Kfig_model}
\end{figure*}

Several works have provided insight into the mechanisms of enhanced conduction in Al-graphene (Al-G) and Cu-graphene (Cu-G) composites \cite{Tokutomi, Chyada, Brown, review_paper_2021, Zhang_2021,kim_2020, Ayar_2021}. For example, Cao and co-workers showed that the electron concentration in both Al and carbon (C) atoms is contingent upon the orientation of the Al-G interface \cite{Cao_2019}. Wang and co-workers demonstrated that the incorporation of graphene additives induces a shift in the Fermi level of copper from \textit{ab initio} calculations \cite{Wang_2011}. These studies suggest that the presence of graphene in aluminum and copper results in the alignment of metal grains in specific orientations and/or facilitates direct carrier transfer between graphene and metals. However, the precise mechanism by which this transport occurs and the impact of graphene on the global conductivity of these composites is still not well understood.

In a previous Letter in this journal \cite{apl_2023}, we discussed electronic transport in copper-graphene composites by considering a single graphene layer sandwiched between two Cu (111) surfaces. We noted the enhancement of the electronic density of states near the Fermi level at short copper-graphene distances - suggesting improved electronic conductivity for the composites. This complementary study extends the analysis to aluminum-graphene composites formed with single and double-layer graphene. We offer atomistic insights into structural relaxation at the aluminum-graphene interface and explore the temperature-dependent conductivity with the number of graphene layers in the composites. In what follows, models of  Aluminum - single-layer graphene - Aluminum composites are referred to as SL, and Aluminum - double-layer graphene - Aluminum composite is referred to as DL. We focus the discussion on the DL models except where the contrast to SL is informative.

To create the Al-G composite models, we started with an orthorhombic cell of face-centered Al (111) that includes a stacking fault, as shown in Figure \ref{fig:Kfig_model} (a). Al (111) terminations are known for their low surface energy and high electronic conductivity \cite{111_orient_1,111_orient_2}. Single and double ("AB" stacking) graphene sheets were positioned above the aluminum fault layer to form an interface. The side view of the arrangement of the atoms in the SL and DL models is shown in Figures \ref{fig:Kfig_model} (b) and \ref{fig:Kfig_model} (c), respectively. To represent the thermophysical phenomena typically observed in solid-phase processed Al-G composites, we simulated the "compression" of the composites by reducing the Al-G distance. A similar method was employed in earlier work to study the pressure dependence of conductivity on mono-crystal Copper, as well as Copper - single-layer Graphene - Copper composite \cite{apl_2023,lanzillo_2014}. Next, employing the conjugate gradient algorithm within the Vienna ab initio simulation package (VASP) \cite{VASP}, an energy-optimized interface structure of the composite was attained for several constant volume simulations. The compression and relaxation procedures employed in this paper hope to mimic conditions commonly observed in aluminum composites that exhibit improved electrical conductivity through solid-phase processing methods. Details regarding the VASP simulation protocol and the method to create the compressed composite models are provided in Sections S1 and S2 of the supplementary material.

Visual inspection of the composite models before any compression and structural optimization, as shown in Figure \ref{fig:Kfig_model} (d), shows the misalignment between Al and C atoms as a honeycomb lattice of graphene (lattice constant of 2.46  \AA{}) has a lattice mismatch of $\approx 5 \%$ with the face-centered Al (111) surface nearest-neighbor distance of 2.34 \AA{}. However, after compression of the models followed by structural optimization (via energy minimization), there appears to be an alignment between the Al and C atoms. Figure \ref{fig:Kfig_model} (e) ([i] and [ii]), corresponding to models with $d_{Al-G} \ = $ 3.41 \AA~and 2.97 \AA~respectively, shows that the extent of atomic alignments between C and Al is dependent on the extent of the compression. The self-organized interface configuration for the compressed model ($d_{Al-G} \ = $ 2.97 \AA{} in this work) is one of the low energy Al-G interface structure, so-called strain-free registry \cite{QI2005155, Woong,PhysRevB.69.235401}.
 
\begin{figure}[!t]
        \centering
	\includegraphics[width=\linewidth]{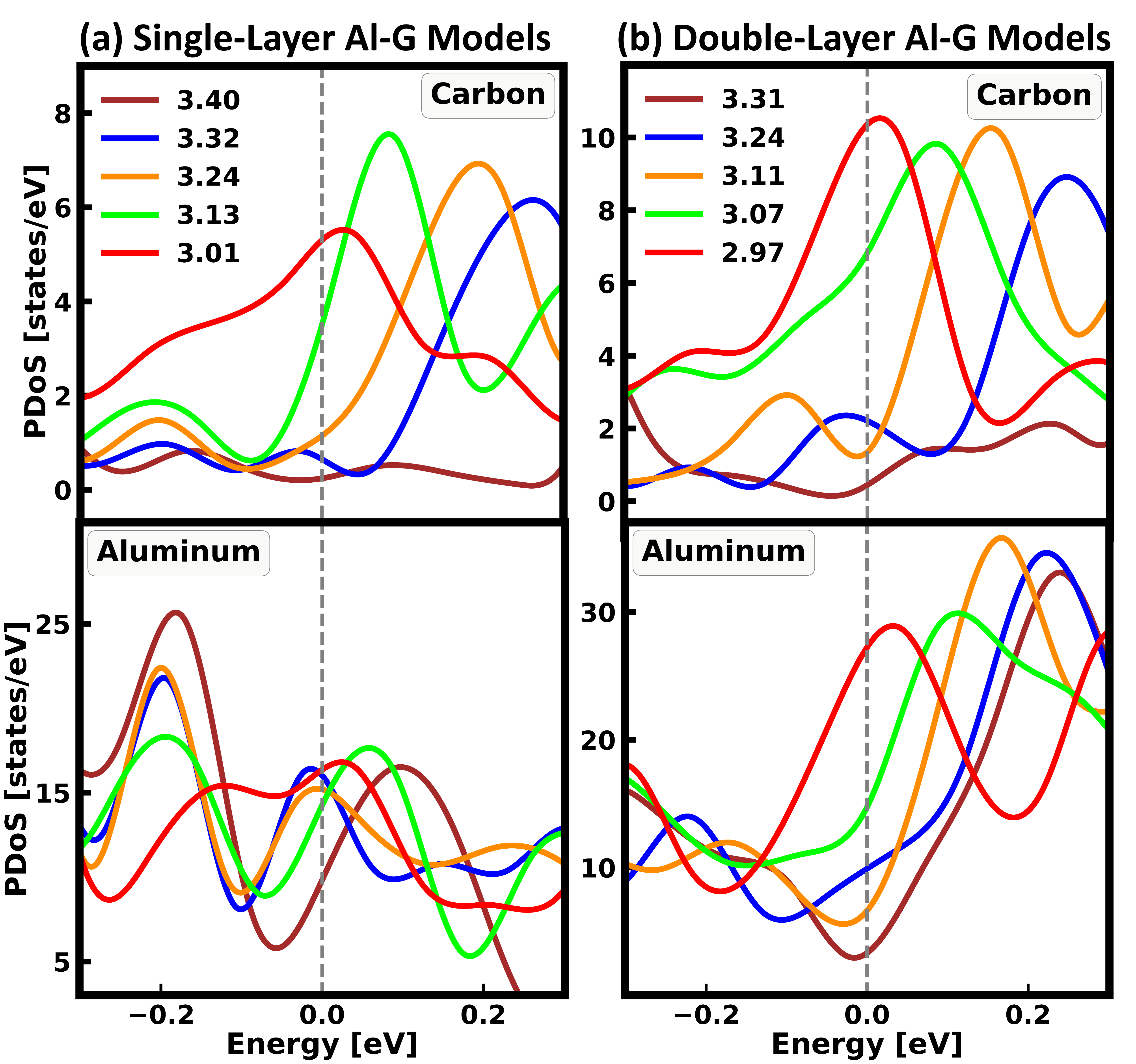}
        \caption{ Projected electron density of states (PDoS) on carbon and aluminum atoms for (a) SL and (b) DL composites. The Fermi level is shifted to zero and is shown by the gray vertical dashed line in each subplot.}
	\label{fig:Kfig_Ef}
\end{figure}

\begin{table}[!t]
         \caption{Fermi level as a function of Al-G composites for varying aluminum-graphene interfacial distance. The first and second row corresponds to DL and SL composites respectvely. }
        \label{tab:Fermi_Level}
		\begin{tabular*}{\linewidth}{@{\extracolsep\fill}c|c|c|c|c|c|c|c|c|c} 
                \hline
              \textbf{$d_{Al-G}$ [\AA]}& 3.41 & 3.35  & 3.31  & 3.24 & 3.19 & 3.11 & 3.07 & 2.97 & 2.94 \\   
              
              \textbf{$E_f$ [eV]} & 6.70& 6.79 & 6.93 & 7.05  & 7.20 & 7.29 & 7.36 & 7.52 & 7.63 \\
             \hline
                \hline
               
              \textbf{$d_{Al-G}$ [\AA]} &3.40 & 3.35  & 3.31  & 3.25 & 3.13 & 3.01 & 2.90 & 2.71 \\ 
              
              \textbf{$E_f$ [eV]} &  6.78 & 7.07 & 7.34 & 7.62 & 7.87 & 8.18 & 8.43 &8.67 \\ 
        
		\end{tabular*}
\end{table}

 \begin{figure*}
	\includegraphics[width=0.97\linewidth]{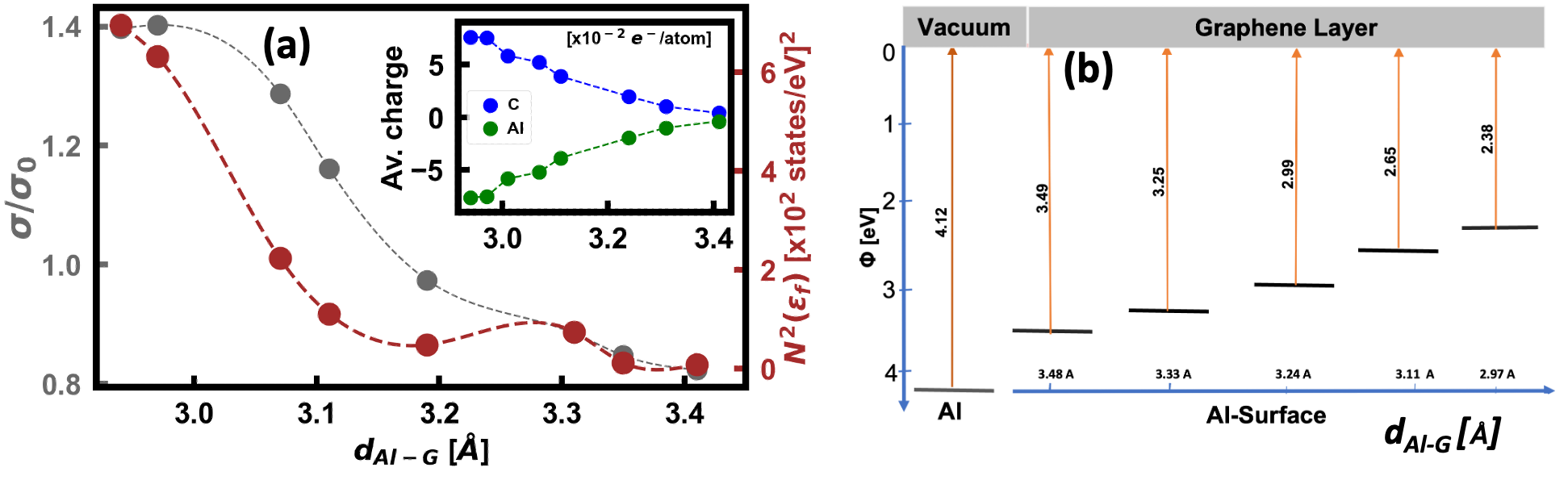}
        \caption{(a) Conductivity for the DL model for various Al-G distances ($d_{Al-G}$) in the x-axis). The average conductivity ($\sigma$) is represented by the gray curve, with $\sigma_{0}$ denoting the conductivity of the Al-matrix shown in Figure \ref{fig:Kfig_model} (a) calculated at 300K. The squared density of states at the Fermi level is shown in brown. In the inset, a Bader analysis \cite{Bader} illustrates the average charge gain and loss for C (blue) and Al (green) atoms. (b) Estimated energy ($\Phi$) required to remove an electron from pure Al surface and with graphene layer placed on it at different interfacial distances. Dotted lines are included in all plots as visual aids.} 
	\label{kfig_DoS_bader}
\end{figure*} 
We computed the atom-projected electronic density of states (PDoS)  for varying interfacial distance models. The PDoS for the SL and DL composite models are shown in Figure \ref{fig:Kfig_Ef} (a) and \ref{fig:Kfig_Ef} (b), respectively. The focus is on the region near the Fermi energy ($\epsilon_f$), indicated by the gray dashed lines and shifted to zero. As the distance between aluminum and graphene (Al-G distance) decreased, we observed an enhancement of the electronic density of states near the Fermi level from both carbon (TOP) and aluminum (BOTTOM) atoms. With a random phase approximation \cite{RPA1,RPA2,RPA3,RPA4}, Mott and Davis showed that the electronic conductivity is proportional to $N^2(\epsilon_f)$ \cite{Mott_davis}, $N(\epsilon_f)$ being the density of states at the Fermi energy ($\epsilon_f$). To explore this further, we plotted the behavior of $N^2(\epsilon_f)$ for the compressed composite models (brown curve in Figure \ref{kfig_DoS_bader} (a). Indeed, $N^2(\epsilon_f)$ roughly tracks the electronic conductivity ($\sigma$) calculated using the Kubo-Greenwood formula (KGF)\cite{Kubo, Greenwood,Mosely&Lukes}, shown by gray curve in Figure \ref{kfig_DoS_bader} (a). As the Al-G distance decreases, both $N^2(\epsilon_f)$ and electronic conductivity increase, consistent with the elementary notion that metallicity/conduction are associated with a large $N(\epsilon_f)$. The most compressed DL composite ($d_{Al-G} =$ 2.97 \AA) exhibits approximately 40\% higher conductivity relative to the aluminum matrix at 300 K. Analogous results for the SL models can be found in Figure S1 in the supplementary material.
 \begin{figure}
     \centering
	\includegraphics[width=\linewidth]{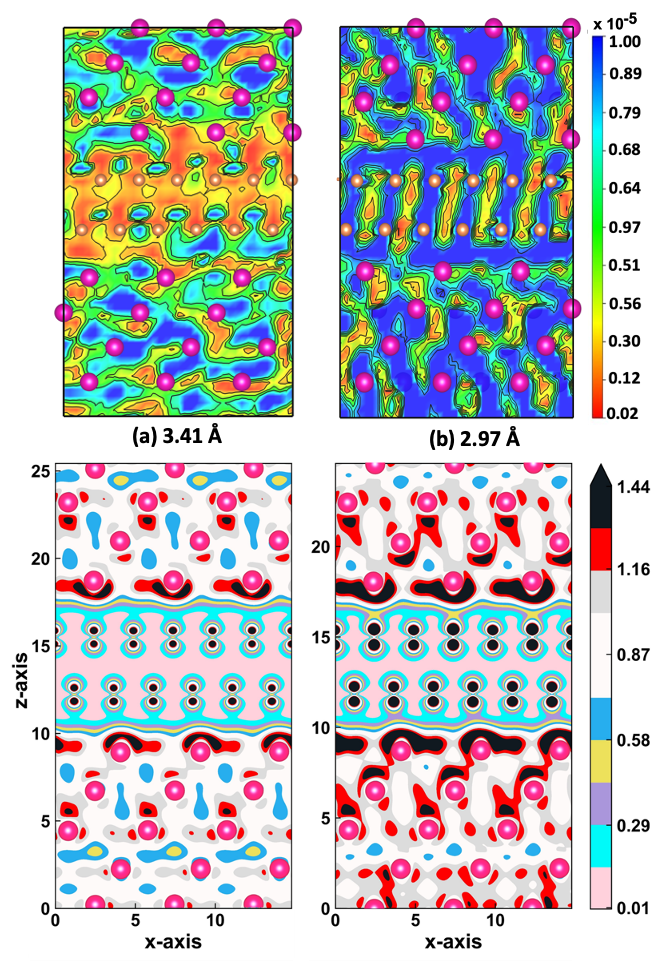}
        \caption{[TOP] Projected 2D transverse SPC [in Siemens/cm/\AA$^{3}$] iso-surface plot. [BOTTOM] Band decomposed charge density [in $e^-$/\AA$^{3}$] corresponds to 15 bands below and above the Fermi level. The charge density values were scaled by a factor of 100. The data presented is for DL models with $d_{Al-G}$ = (a) 3.41 \AA, (b) 2.97 \AA. The pink (brown) spheres represent the Al (C) atoms in the models.}  
	\label{Kfig_KGF_CHGDEN}
\end{figure}

The Fermi level shifts towards higher energies with decreasing Al-G distance, allowing more electronic states to participate in conduction (see details in Table \ref{tab:Fermi_Level}).  This behavior has been reported for graphene on copper \cite{apl_2023,Wang_2011}. We further predicted the work function of the composites for varying Al-G distances which is shown in Figure  \ref{kfig_DoS_bader} (b). The plot shows that the work function decreases with decreasing Al-G distance, and hints at increasing charge transfer between interfacial graphene and aluminum atoms. These effects are quantified by estimating the average charge transfer from interfacial Al to C atoms, shown in the inset of Figure \ref{kfig_DoS_bader} (a). At the shortest interfacial distance ($d_{Al-G} = $ 2.97 \AA), the average transfer of electronic charge to graphene reached 0.075 electrons per atom.

Next, we computed the conduction path in real space and its Al-G distance dependence. To achieve this, we employed the space-projected conductivity (SPC) method to project the electronic conductivity onto real-space grids \cite{K1,K2,K4}. The upper panel of Figure \ref{Kfig_KGF_CHGDEN} shows the isosurface plots of the transverse SPC values for DL composite models with interfacial distances of 3.41 \AA~ and 2.97 \AA, represented by (a) and (b) in Figure \ref{Kfig_KGF_CHGDEN}, respectively. The color bar on the right indicates the magnitude of SPC values, with red (blue) indicating low (high) values. At short Al-G distances, both aluminum and graphene contribute to conduction, particularly at the Al-G interface. The SPC at short Al-G distances reveals that graphene actively participates in conduction and forms a bridge between Al atoms on opposite layers. Notably, Figure \ref{Kfig_KGF_CHGDEN} (upper panel) illustrates the formation of a continuous network of graphene sheets within the aluminum matrix, establishing a pathway for electron transport.

To further delineate the enhanced electron transport through the Al-G interface, we computed the electronic charge density near the Fermi level (see implementation examples in References\cite{aF,nt,aG2,prl_raj}). By decomposing 15 bands above and below the Fermi level from the total electron charge density, we generated isosurface plots, shown in the lower panel of Figure \ref{Kfig_KGF_CHGDEN}. The same models used for the SPC calculation were employed. In the model with $d_{Al-G} = 2.97$ \AA{}, a higher degree of interaction between graphene and interfacial aluminum atoms was observed, indicated by the presence of black and red regions in the isosurface plot. 

Next, we investigated the temperature dependence of conductivity in the composite models. This was done by estimating the average electrical conductivity from KGF for models held at different temperatures. The procedure to calculate the temperature-dependent conductivity is discussed in Section S3 of the supplementary material. Figure \ref{fig:Kfig_Temp} presents the average electronic conductivity, obtained from 10 uncorrelated snapshots, as a function of temperature, ranging from 100 K to 600 K. Figure \ref{fig:Kfig_Temp} (a) shows the conductivity behavior for the DL models corresponding to two interfacial distances. The model with $d_{Al-G}$ = 3.41 \AA~ exhibits a nearly linear relationship, shown by blue plots. However, the model with short Al-G distance, $d_{Al-G}$ = 2.97 \AA~ (shown by red plots), displays local extrema at around 300 K (minima) and 400 K (maxima). This non-monotonic behavior is in accord with experimental observations of conductivity enhancement in solid-phase processed metal-graphene composites \cite{Aditya_MSE,Gwalani_MD}, suggesting that this work captures, to some extent, the physics of the real material. In contrast, the extrema are not observed in the SL composite models, as shown in Figure \ref{fig:Kfig_Temp} (b) and \ref{fig:Kfig_Temp} (c) which correspond to Al-G distances $d_{Al-G}$ = 3.01 \AA{} and 3.40 \AA{} respectively.  The non-monotonic temperature dependence observed exclusively in the compressed DL composites can be attributed to two factors: (1) The active involvement of graphene layers in charge transfer at shorter Al-G distances and (2) thermally driven hopping across the inter-layer galleries between the compressed graphene double-layer \cite{ONO, Norio_graphite, Bapat_carbon, matsubara}.

\begin{figure}
\centering
    \includegraphics[width=\linewidth]{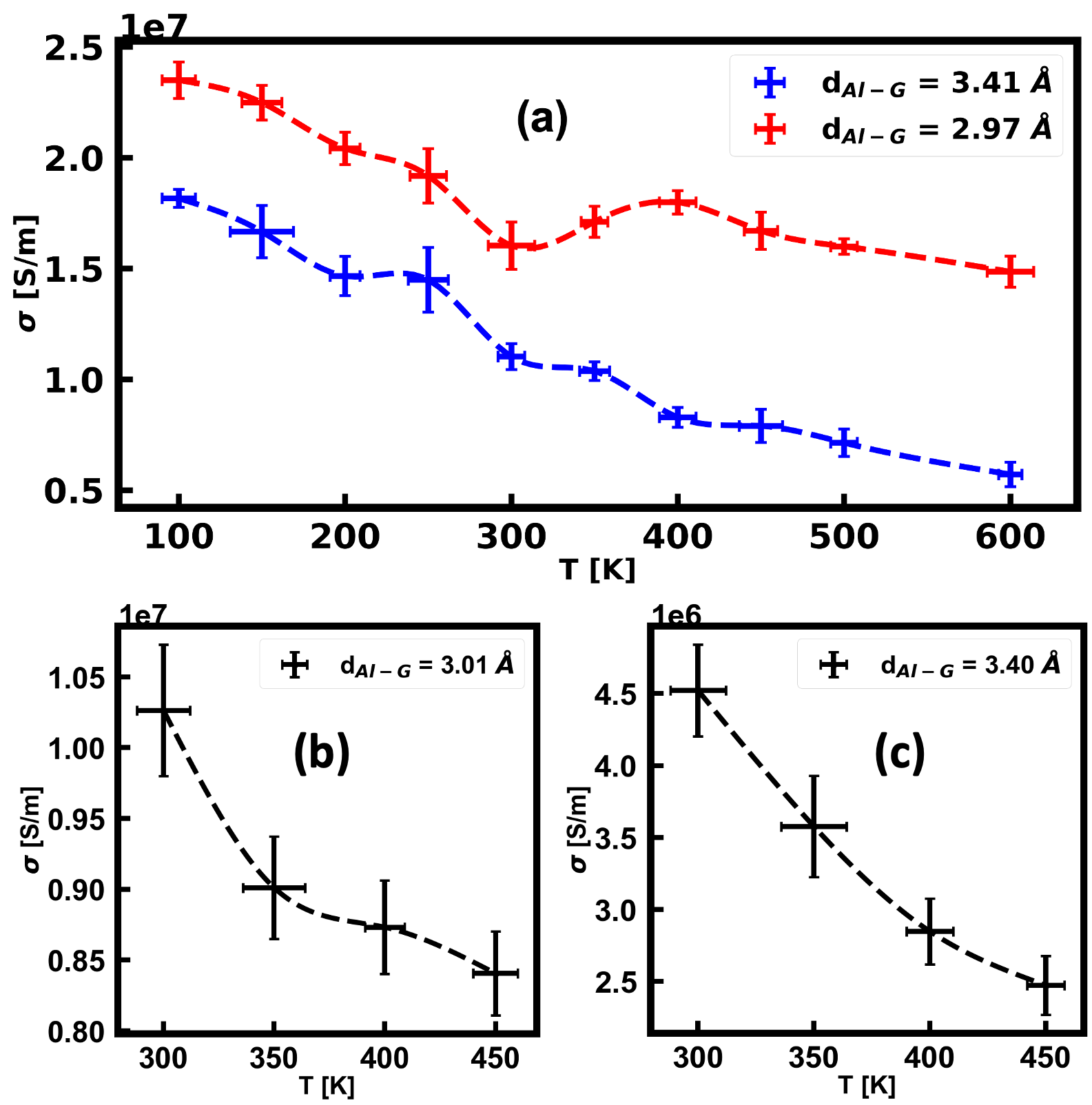}
        \caption{(a) Average electronic conductivity plotted versus annealing temperature for DL composite models with Al-G distances of 2.97 \AA~(red) and 3.41 \AA~(blue). (b) and (c) Similar plots for SL composite models, with Al-G distances of 3.01 \AA~and 3.40 \AA, respectively. Vertical bars represent the standard deviation from the mean conductivity, averaged from the last 10 snapshots taken at 50 fs intervals over 3 ps annealing. Horizontal bars represent temperature fluctuations during constant temperature annealing.}
 \label{fig:Kfig_Temp}
\end{figure}
 
We show that graphene and graphene stacks enhance the electronic conductivity of Al and while a key addition to the area, it is not the full story. The graphene structures are dispersed in an unknown way throughout the metal microstructures in the experimentally synthesized bulk composites and conspire to create a globally enhanced conductivity and globally modified temperature dependence. These effects could be due to (1) reduced scattering at grain boundaries from the graphene or (2) forming a network of isolated or weakly interacting Al-G structures. Our work complements both of these imaginings. The registry between the $sp^2$ carbon network (see Figure \ref{fig:Kfig_model}e [ii])  and the Al (111) surface suggests that mechanism (1) may be a key player in conductivity enhancement, as we demonstrate that a self-organized grain boundary buffer may form at Al (111) surfaces.

In conclusion, this study provides a comprehensive atomic-level understanding of the role of graphene as an additive in aluminum grains, focusing on single- and double-graphene stack(s) in the aluminum matrix. We have demonstrated that the increased electrical conductivity observed in Al-graphene composites arises from the enhanced electronic dynamics at the Fermi level. The interaction between carbon and interfacial aluminum atoms highlights the active role of graphene in facilitating electronic conduction. Furthermore, our study depicts the experimentally observed enhanced electrical conductivity within the temperature range of 300 K to 400 K.

\begin{acknowledgments}
The authors gratefully acknowledge the support received from the National Science Foundation (NSF) for computational resources through XSEDE (Grant No. ACI-1548562; allocation no. DMR-190008P) and ACCESS (Grant No. 2138259, 2138286, and 2138296; allocation no. phy230007p).  The authors also acknowledge the support received from the Department of Energy (DOE) Vehicles Technology Office Powertrain Materials Core Program.  Pacific Northwest National Laboratory is operated by the Battelle Memorial Institute for the U.S. Department of Energy under contract No. DE-AC06-76LO1830. 
\end{acknowledgments}


\bibliography{APL_AL_G_biblography}

\providecommand{\noopsort}[1]{}\providecommand{\singleletter}[1]{#1}%
\begin{thebibliography}{43}%
\makeatletter
\providecommand \@ifxundefined [1]{%
 \@ifx{#1\undefined}
}%
\providecommand \@ifnum [1]{%
 \ifnum #1\expandafter \@firstoftwo
 \else \expandafter \@secondoftwo
 \fi
}%
\providecommand \@ifx [1]{%
 \ifx #1\expandafter \@firstoftwo
 \else \expandafter \@secondoftwo
 \fi
}%
\providecommand \natexlab [1]{#1}%
\providecommand \enquote  [1]{``#1''}%
\providecommand \bibnamefont  [1]{#1}%
\providecommand \bibfnamefont [1]{#1}%
\providecommand \citenamefont [1]{#1}%
\providecommand \href@noop [0]{\@secondoftwo}%
\providecommand \href [0]{\begingroup \@sanitize@url \@href}%
\providecommand \@href[1]{\@@startlink{#1}\@@href}%
\providecommand \@@href[1]{\endgroup#1\@@endlink}%
\providecommand \@sanitize@url [0]{\catcode `\\12\catcode `\$12\catcode `\&12\catcode `\#12\catcode `\^12\catcode `\_12\catcode `\%12\relax}%
\providecommand \@@startlink[1]{}%
\providecommand \@@endlink[0]{}%
\providecommand \url  [0]{\begingroup\@sanitize@url \@url }%
\providecommand \@url [1]{\endgroup\@href {#1}{\urlprefix }}%
\providecommand \urlprefix  [0]{URL }%
\providecommand \Eprint [0]{\href }%
\providecommand \doibase [0]{http://dx.doi.org/}%
\providecommand \selectlanguage [0]{\@gobble}%
\providecommand \bibinfo  [0]{\@secondoftwo}%
\providecommand \bibfield  [0]{\@secondoftwo}%
\providecommand \translation [1]{[#1]}%
\providecommand \BibitemOpen [0]{}%
\providecommand \bibitemStop [0]{}%
\providecommand \bibitemNoStop [0]{.\EOS\space}%
\providecommand \EOS [0]{\spacefactor3000\relax}%
\providecommand \BibitemShut  [1]{\csname bibitem#1\endcsname}%
\let\auto@bib@innerbib\@empty
\bibitem [{\citenamefont {Nittala}\ \emph {et~al.}(2023)\citenamefont {Nittala}, \citenamefont {Smith}, \citenamefont {Gwalani}, \citenamefont {Silverstein}, \citenamefont {Kraft},\ and\ \citenamefont {Kappagantula}}]{Aditya_MSE}%
  \BibitemOpen
  \bibfield  {author} {\bibinfo {author} {\bibfnamefont {A.}~\bibnamefont {Nittala}}, \bibinfo {author} {\bibfnamefont {J.}~\bibnamefont {Smith}}, \bibinfo {author} {\bibfnamefont {B.}~\bibnamefont {Gwalani}}, \bibinfo {author} {\bibfnamefont {J.}~\bibnamefont {Silverstein}}, \bibinfo {author} {\bibfnamefont {F.}~\bibnamefont {Kraft}}, \ and\ \bibinfo {author} {\bibfnamefont {K.}~\bibnamefont {Kappagantula}},\ }\bibfield  {title} {\enquote {\bibinfo {title} {Simultaneously improved electrical and mechanical performance of hot-extruded bulk scale aluminum-graphene wires},}\ }\href {\doibase 10.1016/j.mseb.2023.116452} {\bibfield  {journal} {\bibinfo  {journal} {Materials Science and Engineering: B}\ }\textbf {\bibinfo {volume} {293}},\ \bibinfo {pages} {116452} (\bibinfo {year} {2023})}\BibitemShut {NoStop}%
\bibitem [{\citenamefont {Kappagantula}\ \emph {et~al.}(2022)\citenamefont {Kappagantula}, \citenamefont {Smith}, \citenamefont {Nittala},\ and\ \citenamefont {Kraft}}]{KAPPAGANTULA2022162477}%
  \BibitemOpen
  \bibfield  {author} {\bibinfo {author} {\bibfnamefont {K.~S.}\ \bibnamefont {Kappagantula}}, \bibinfo {author} {\bibfnamefont {J.~A.}\ \bibnamefont {Smith}}, \bibinfo {author} {\bibfnamefont {A.~K.}\ \bibnamefont {Nittala}}, \ and\ \bibinfo {author} {\bibfnamefont {F.~F.}\ \bibnamefont {Kraft}},\ }\bibfield  {title} {\enquote {\bibinfo {title} {Macro copper-graphene composites with enhanced electrical conductivity},}\ }\href {\doibase https://doi.org/10.1016/j.jallcom.2021.162477} {\bibfield  {journal} {\bibinfo  {journal} {Journal of Alloys and Compounds}\ }\textbf {\bibinfo {volume} {894}},\ \bibinfo {pages} {162477} (\bibinfo {year} {2022})}\BibitemShut {NoStop}%
\bibitem [{\citenamefont {Gwalani}\ \emph {et~al.}(2023)\citenamefont {Gwalani}, \citenamefont {Li}, \citenamefont {Nittala}, \citenamefont {Choi}, \citenamefont {Reza-E-Rabby}, \citenamefont {Atehortua}, \citenamefont {Bhattacharjee}, \citenamefont {Pole}, \citenamefont {Silverstein}, \citenamefont {Song},\ and\ \citenamefont {Kappagantula}}]{Gwalani_MD}%
  \BibitemOpen
  \bibfield  {author} {\bibinfo {author} {\bibfnamefont {B.}~\bibnamefont {Gwalani}}, \bibinfo {author} {\bibfnamefont {X.}~\bibnamefont {Li}}, \bibinfo {author} {\bibfnamefont {A.}~\bibnamefont {Nittala}}, \bibinfo {author} {\bibfnamefont {W.}~\bibnamefont {Choi}}, \bibinfo {author} {\bibfnamefont {M.}~\bibnamefont {Reza-E-Rabby}}, \bibinfo {author} {\bibfnamefont {J.}~\bibnamefont {Atehortua}}, \bibinfo {author} {\bibfnamefont {A.}~\bibnamefont {Bhattacharjee}}, \bibinfo {author} {\bibfnamefont {M.}~\bibnamefont {Pole}}, \bibinfo {author} {\bibfnamefont {J.}~\bibnamefont {Silverstein}}, \bibinfo {author} {\bibfnamefont {M.}~\bibnamefont {Song}}, \ and\ \bibinfo {author} {\bibfnamefont {K.}~\bibnamefont {Kappagantula}},\ }\bibfield  {title} {\enquote {\bibinfo {title} {Unprecedented electrical performance of friction-extruded copper-graphene composites},}\ }\href {\doibase 10.1016/j.matdes.2023.112555} {\bibfield  {journal} {\bibinfo  {journal} {Materials and Design}\ }\textbf {\bibinfo {volume} {237}},\
  \bibinfo {pages} {112555} (\bibinfo {year} {2023})}\BibitemShut {NoStop}%
\bibitem [{\citenamefont {Sharma}, \citenamefont {Bhandari},\ and\ \citenamefont {Pinca-Bretotean}(2021)}]{SHARMA20214133}%
  \BibitemOpen
  \bibfield  {author} {\bibinfo {author} {\bibfnamefont {A.~K.}\ \bibnamefont {Sharma}}, \bibinfo {author} {\bibfnamefont {R.}~\bibnamefont {Bhandari}}, \ and\ \bibinfo {author} {\bibfnamefont {C.}~\bibnamefont {Pinca-Bretotean}},\ }\bibfield  {title} {\enquote {\bibinfo {title} {A systematic overview on fabrication aspects and methods of aluminum metal matrix composites},}\ }\href {\doibase https://doi.org/10.1016/j.matpr.2020.11.899} {\bibfield  {journal} {\bibinfo  {journal} {Materials Today: Proceedings}\ }\textbf {\bibinfo {volume} {45}},\ \bibinfo {pages} {4133--4138} (\bibinfo {year} {2021})},\ \bibinfo {note} {8th International Conference on Advanced Materials and Structures - AMS 2020}\BibitemShut {NoStop}%
\bibitem [{\citenamefont {Jiju}, \citenamefont {Gurusamy},\ and\ \citenamefont {Prakash}(2020)}]{article2}%
  \BibitemOpen
  \bibfield  {author} {\bibinfo {author} {\bibfnamefont {K.}~\bibnamefont {Jiju}}, \bibinfo {author} {\bibfnamefont {S.}~\bibnamefont {Gurusamy}}, \ and\ \bibinfo {author} {\bibfnamefont {S.}~\bibnamefont {Prakash}},\ }\bibfield  {title} {\enquote {\bibinfo {title} {Study on preparation of al – sic metal matrix composites using powder metallurgy technique and its mechanical properties},}\ }\href {\doibase https://doi.org/10.1016/j.matpr.2020.04.001} {\bibfield  {journal} {\bibinfo  {journal} {Materials Today: Proceedings}\ }\textbf {\bibinfo {volume} {27}},\ \bibinfo {pages} {1843--1847} (\bibinfo {year} {2020})}\BibitemShut {NoStop}%
\bibitem [{\citenamefont {Sauvage}\ \emph {et~al.}(2015)\citenamefont {Sauvage}, \citenamefont {Bobruk}, \citenamefont {Murashkin}, \citenamefont {Nasedkina}, \citenamefont {Enikeev},\ and\ \citenamefont {Valiev}}]{Sauvage_2015}%
  \BibitemOpen
  \bibfield  {author} {\bibinfo {author} {\bibfnamefont {X.}~\bibnamefont {Sauvage}}, \bibinfo {author} {\bibfnamefont {E.~V.}\ \bibnamefont {Bobruk}}, \bibinfo {author} {\bibfnamefont {M.~Y.}\ \bibnamefont {Murashkin}}, \bibinfo {author} {\bibfnamefont {Y.}~\bibnamefont {Nasedkina}}, \bibinfo {author} {\bibfnamefont {N.~A.}\ \bibnamefont {Enikeev}}, \ and\ \bibinfo {author} {\bibfnamefont {R.~Z.}\ \bibnamefont {Valiev}},\ }\bibfield  {title} {\enquote {\bibinfo {title} {Optimization of electrical conductivity and strength combination by structure design at the nanoscale in al–mg–si alloys},}\ }\href {\doibase https://doi.org/10.1016/j.actamat.2015.07.039} {\bibfield  {journal} {\bibinfo  {journal} {Acta Mater.}\ }\textbf {\bibinfo {volume} {98}},\ \bibinfo {pages} {355} (\bibinfo {year} {2015})}\BibitemShut {NoStop}%
\bibitem [{\citenamefont {Tokutomi}\ \emph {et~al.}(2015)\citenamefont {Tokutomi}, \citenamefont {Uemura}, \citenamefont {Sugiyama}, \citenamefont {Shiomi},\ and\ \citenamefont {Yanagimoto}}]{Tokutomi}%
  \BibitemOpen
  \bibfield  {author} {\bibinfo {author} {\bibfnamefont {J.}~\bibnamefont {Tokutomi}}, \bibinfo {author} {\bibfnamefont {T.}~\bibnamefont {Uemura}}, \bibinfo {author} {\bibfnamefont {S.}~\bibnamefont {Sugiyama}}, \bibinfo {author} {\bibfnamefont {J.}~\bibnamefont {Shiomi}}, \ and\ \bibinfo {author} {\bibfnamefont {J.}~\bibnamefont {Yanagimoto}},\ }\bibfield  {title} {\enquote {\bibinfo {title} {Hot extrusion to manufacture the metal matrix composite of carbon nanotube and aluminum with excellent electrical conductivities and mechanical properties.}}\ }\href {\doibase 10.1016/j.cirp.2015.04.083} {\bibfield  {journal} {\bibinfo  {journal} {CIRP Annals - Manufacturing Technology}\ }\textbf {\bibinfo {volume} {64}},\ \bibinfo {pages} {257--260} (\bibinfo {year} {2015})}\BibitemShut {NoStop}%
\bibitem [{\citenamefont {Chyada}, \citenamefont {Jabur},\ and\ \citenamefont {Alwan}(2017)}]{Chyada}%
  \BibitemOpen
  \bibfield  {author} {\bibinfo {author} {\bibfnamefont {F.~A.}\ \bibnamefont {Chyada}}, \bibinfo {author} {\bibfnamefont {A.~R.}\ \bibnamefont {Jabur}}, \ and\ \bibinfo {author} {\bibfnamefont {H.~A.}\ \bibnamefont {Alwan}},\ }\bibfield  {title} {\enquote {\bibinfo {title} {Effect addition of graphene on electrical conductivity and tensile strength for recycled electric power transmission wires},}\ }\href {\doibase https://doi.org/10.1016/j.egypro.2017.07.055} {\bibfield  {journal} {\bibinfo  {journal} {Energy Procedia}\ }\textbf {\bibinfo {volume} {119}},\ \bibinfo {pages} {121–130} (\bibinfo {year} {2017})}\BibitemShut {NoStop}%
\bibitem [{\citenamefont {Brown}\ \emph {et~al.}(2014)\citenamefont {Brown}, \citenamefont {Joyce}, \citenamefont {Forrest},\ and\ \citenamefont {Salamanca-Riba}}]{Brown}%
  \BibitemOpen
  \bibfield  {author} {\bibinfo {author} {\bibfnamefont {L.}~\bibnamefont {Brown}}, \bibinfo {author} {\bibfnamefont {P.}~\bibnamefont {Joyce}}, \bibinfo {author} {\bibfnamefont {D.}~\bibnamefont {Forrest}}, \ and\ \bibinfo {author} {\bibfnamefont {L.}~\bibnamefont {Salamanca-Riba}},\ }\bibfield  {title} {\enquote {\bibinfo {title} {Physical and mechanical characterization of a nanocarbon infused aluminum-matrix composite},}\ }\href {\doibase https://doi.org/10.1520/MPC20130023} {\bibfield  {journal} {\bibinfo  {journal} {Materials Performance and Characterization}\ }\textbf {\bibinfo {volume} {3}},\ \bibinfo {pages} {65--80} (\bibinfo {year} {2014})}\BibitemShut {NoStop}%
\bibitem [{\citenamefont {Ali}\ \emph {et~al.}(2021)\citenamefont {Ali}, \citenamefont {Omar}, \citenamefont {Hashim}, \citenamefont {Salleh},\ and\ \citenamefont {Mohamed}}]{review_paper_2021}%
  \BibitemOpen
  \bibfield  {author} {\bibinfo {author} {\bibfnamefont {A.~M.}\ \bibnamefont {Ali}}, \bibinfo {author} {\bibfnamefont {M.~Z.}\ \bibnamefont {Omar}}, \bibinfo {author} {\bibfnamefont {H.}~\bibnamefont {Hashim}}, \bibinfo {author} {\bibfnamefont {M.~S.}\ \bibnamefont {Salleh}}, \ and\ \bibinfo {author} {\bibfnamefont {I.~F.}\ \bibnamefont {Mohamed}},\ }\bibfield  {title} {\enquote {\bibinfo {title} {Recent development in graphene-reinforced aluminium matrix composite: A review},}\ }\href {\doibase doi:10.1515/rams-2021-0062} {\bibfield  {journal} {\bibinfo  {journal} {Reviews on Advanced Materials Science}\ }\textbf {\bibinfo {volume} {60}},\ \bibinfo {pages} {801--817} (\bibinfo {year} {2021})}\BibitemShut {NoStop}%
\bibitem [{\citenamefont {Zhang}\ and\ \citenamefont {Wang}(2021)}]{Zhang_2021}%
  \BibitemOpen
  \bibfield  {author} {\bibinfo {author} {\bibfnamefont {X.}~\bibnamefont {Zhang}}\ and\ \bibinfo {author} {\bibfnamefont {S.}~\bibnamefont {Wang}},\ }\bibfield  {title} {\enquote {\bibinfo {title} {Interfacial strengthening of graphene/aluminum composites through point defects: A first-principles study.}}\ }\href {\doibase doi: 10.3390/nano11030738} {\bibfield  {journal} {\bibinfo  {journal} {Nanomaterials (Basel)}\ }\textbf {\bibinfo {volume} {11}},\ \bibinfo {pages} {3} (\bibinfo {year} {2021})}\BibitemShut {NoStop}%
\bibitem [{\citenamefont {Kim}\ and\ \citenamefont {Choi}(2020)}]{kim_2020}%
  \BibitemOpen
  \bibfield  {author} {\bibinfo {author} {\bibfnamefont {D.-Y.}\ \bibnamefont {Kim}}\ and\ \bibinfo {author} {\bibfnamefont {H.-J.}\ \bibnamefont {Choi}},\ }\bibfield  {title} {\enquote {\bibinfo {title} {Recent developments towards commercialization of metal matrix composites},}\ }\href {\doibase https://doi.org/10.3390/ma13122828} {\bibfield  {journal} {\bibinfo  {journal} {Materials}\ }\textbf {\bibinfo {volume} {13}},\ \bibinfo {pages} {2828} (\bibinfo {year} {2020})}\BibitemShut {NoStop}%
\bibitem [{\citenamefont {Ayar}, \citenamefont {George},\ and\ \citenamefont {Patel}(2021)}]{Ayar_2021}%
  \BibitemOpen
  \bibfield  {author} {\bibinfo {author} {\bibfnamefont {M.~S.}\ \bibnamefont {Ayar}}, \bibinfo {author} {\bibfnamefont {P.~M.}\ \bibnamefont {George}}, \ and\ \bibinfo {author} {\bibfnamefont {R.~R.}\ \bibnamefont {Patel}},\ }\bibfield  {title} {\enquote {\bibinfo {title} {Advanced research progresses in aluminium metal matrix composites: An overview},}\ }\href {\doibase 10.1063/5.0036141} {\bibfield  {journal} {\bibinfo  {journal} {AIP Conference Proceedings}\ }\textbf {\bibinfo {volume} {2317}},\ \bibinfo {pages} {020026} (\bibinfo {year} {2021})}\BibitemShut {NoStop}%
\bibitem [{\citenamefont {Cao}\ \emph {et~al.}(2019)\citenamefont {Cao}, \citenamefont {Luo}, \citenamefont {Xie}, \citenamefont {Tan}, \citenamefont {Fan}, \citenamefont {Guo}, \citenamefont {Su}, \citenamefont {Li},\ and\ \citenamefont {Xiong}}]{Cao_2019}%
  \BibitemOpen
  \bibfield  {author} {\bibinfo {author} {\bibfnamefont {M.}~\bibnamefont {Cao}}, \bibinfo {author} {\bibfnamefont {Y.}~\bibnamefont {Luo}}, \bibinfo {author} {\bibfnamefont {Y.}~\bibnamefont {Xie}}, \bibinfo {author} {\bibfnamefont {Z.}~\bibnamefont {Tan}}, \bibinfo {author} {\bibfnamefont {G.}~\bibnamefont {Fan}}, \bibinfo {author} {\bibfnamefont {Q.}~\bibnamefont {Guo}}, \bibinfo {author} {\bibfnamefont {Y.}~\bibnamefont {Su}}, \bibinfo {author} {\bibfnamefont {Z.}~\bibnamefont {Li}}, \ and\ \bibinfo {author} {\bibfnamefont {D.-B.}\ \bibnamefont {Xiong}},\ }\bibfield  {title} {\enquote {\bibinfo {title} {The influence of interface structure on the electrical conductivity of graphene embedded in aluminum matrix},}\ }\href {\doibase https://doi.org/10.1002/admi.201900468} {\bibfield  {journal} {\bibinfo  {journal} {Advanced Materials Interfaces}\ }\textbf {\bibinfo {volume} {6}},\ \bibinfo {pages} {1900468} (\bibinfo {year} {2019})}\BibitemShut {NoStop}%
\bibitem [{\citenamefont {Wang}\ \emph {et~al.}(2011)\citenamefont {Wang}, \citenamefont {Liu}, \citenamefont {Wang}, \citenamefont {Sheng},\ and\ \citenamefont {Yu}}]{Wang_2011}%
  \BibitemOpen
  \bibfield  {author} {\bibinfo {author} {\bibfnamefont {W.}~\bibnamefont {Wang}}, \bibinfo {author} {\bibfnamefont {Y.}~\bibnamefont {Liu}}, \bibinfo {author} {\bibfnamefont {T.}~\bibnamefont {Wang}}, \bibinfo {author} {\bibfnamefont {K.}~\bibnamefont {Sheng}}, \ and\ \bibinfo {author} {\bibfnamefont {B.}~\bibnamefont {Yu}},\ }\bibfield  {title} {\enquote {\bibinfo {title} {Graphene/cu (111) interface study: The density functional theory calculations},}\ }in\ \href {\doibase https://doi.org/10.1109/ICECC.2011.6067884} {\emph {\bibinfo {booktitle} {2011 International Conference on Electronics, Communications and Control (ICECC)}}}\ (\bibinfo {organization} {IEEE},\ \bibinfo {year} {2011})\ pp.\ \bibinfo {pages} {265--268}\BibitemShut {NoStop}%
\bibitem [{\citenamefont {Subedi}\ \emph {et~al.}(2023)\citenamefont {Subedi}, \citenamefont {Nepal}, \citenamefont {Ugwumadu}, \citenamefont {Kappagantula},\ and\ \citenamefont {Drabold}}]{apl_2023}%
  \BibitemOpen
  \bibfield  {author} {\bibinfo {author} {\bibfnamefont {K.~N.}\ \bibnamefont {Subedi}}, \bibinfo {author} {\bibfnamefont {K.}~\bibnamefont {Nepal}}, \bibinfo {author} {\bibfnamefont {C.}~\bibnamefont {Ugwumadu}}, \bibinfo {author} {\bibfnamefont {K.}~\bibnamefont {Kappagantula}}, \ and\ \bibinfo {author} {\bibfnamefont {D.~A.}\ \bibnamefont {Drabold}},\ }\bibfield  {title} {\enquote {\bibinfo {title} {Electronic transport in copper–graphene composites},}\ }\href {\doibase https://doi.org/10.1063/5.0137086} {\bibfield  {journal} {\bibinfo  {journal} {Applied Physics Letters}\ }\textbf {\bibinfo {volume} {122}},\ \bibinfo {pages} {031903} (\bibinfo {year} {2023})}\BibitemShut {NoStop}%
\bibitem [{\citenamefont {Zhang}, \citenamefont {Ma},\ and\ \citenamefont {Xu}(2004)}]{111_orient_1}%
  \BibitemOpen
  \bibfield  {author} {\bibinfo {author} {\bibfnamefont {J.-M.}\ \bibnamefont {Zhang}}, \bibinfo {author} {\bibfnamefont {F.}~\bibnamefont {Ma}}, \ and\ \bibinfo {author} {\bibfnamefont {K.-W.}\ \bibnamefont {Xu}},\ }\bibfield  {title} {\enquote {\bibinfo {title} {Calculation of the surface energy of fcc metals with modified embedded-atom method},}\ }\href {\doibase https://doi.org/10.1016/j.apsusc.2003.09.050} {\bibfield  {journal} {\bibinfo  {journal} {Applied Surface Science}\ }\textbf {\bibinfo {volume} {229}},\ \bibinfo {pages} {34--42} (\bibinfo {year} {2004})}\BibitemShut {NoStop}%
\bibitem [{\citenamefont {Wang}\ \emph {et~al.}(2022)\citenamefont {Wang}, \citenamefont {Li}, \citenamefont {Peng}, \citenamefont {Gao}, \citenamefont {Wang},\ and\ \citenamefont {Sun}}]{111_orient_2}%
  \BibitemOpen
  \bibfield  {author} {\bibinfo {author} {\bibfnamefont {Y.}~\bibnamefont {Wang}}, \bibinfo {author} {\bibfnamefont {M.}~\bibnamefont {Li}}, \bibinfo {author} {\bibfnamefont {P.}~\bibnamefont {Peng}}, \bibinfo {author} {\bibfnamefont {H.}~\bibnamefont {Gao}}, \bibinfo {author} {\bibfnamefont {J.}~\bibnamefont {Wang}}, \ and\ \bibinfo {author} {\bibfnamefont {B.}~\bibnamefont {Sun}},\ }\bibfield  {title} {\enquote {\bibinfo {title} {Preferred orientation at the al/graphene interface: First-principles calculations and experimental observation},}\ }\href {\doibase https://doi.org/10.1016/j.jallcom.2021.163304} {\bibfield  {journal} {\bibinfo  {journal} {Journal of Alloys and Compounds}\ }\textbf {\bibinfo {volume} {900}},\ \bibinfo {pages} {163304} (\bibinfo {year} {2022})}\BibitemShut {NoStop}%
\bibitem [{\citenamefont {Lanzillo}\ \emph {et~al.}(2014)\citenamefont {Lanzillo}, \citenamefont {Thomas}, \citenamefont {Watson}, \citenamefont {Washington},\ and\ \citenamefont {Nayak}}]{lanzillo_2014}%
  \BibitemOpen
  \bibfield  {author} {\bibinfo {author} {\bibfnamefont {N.~A.}\ \bibnamefont {Lanzillo}}, \bibinfo {author} {\bibfnamefont {J.~B.}\ \bibnamefont {Thomas}}, \bibinfo {author} {\bibfnamefont {B.}~\bibnamefont {Watson}}, \bibinfo {author} {\bibfnamefont {M.}~\bibnamefont {Washington}}, \ and\ \bibinfo {author} {\bibfnamefont {S.~K.}\ \bibnamefont {Nayak}},\ }\bibfield  {title} {\enquote {\bibinfo {title} {Pressure-enabled phonon engineering in metals},}\ }\href {\doibase 10.1073/pnas.1406721111} {\bibfield  {journal} {\bibinfo  {journal} {Proceedings of the National Academy of Sciences}\ }\textbf {\bibinfo {volume} {111}},\ \bibinfo {pages} {8712--8716} (\bibinfo {year} {2014})}\BibitemShut {NoStop}%
\bibitem [{\citenamefont {Kresse}\ and\ \citenamefont {Hafner}(1993)}]{VASP}%
  \BibitemOpen
  \bibfield  {author} {\bibinfo {author} {\bibfnamefont {G.}~\bibnamefont {Kresse}}\ and\ \bibinfo {author} {\bibfnamefont {J.}~\bibnamefont {Hafner}},\ }\bibfield  {title} {\enquote {\bibinfo {title} {Ab initio molecular dynamics for liquid metals},}\ }\href {\doibase 10.1103/PhysRevB.47.558} {\bibfield  {journal} {\bibinfo  {journal} {Phys. Rev. B}\ }\textbf {\bibinfo {volume} {47}},\ \bibinfo {pages} {558--561} (\bibinfo {year} {1993})}\BibitemShut {NoStop}%
\bibitem [{\citenamefont {Qi}\ \emph {et~al.}(2005)\citenamefont {Qi}, \citenamefont {Hector}, \citenamefont {Ooi},\ and\ \citenamefont {Adams}}]{QI2005155}%
  \BibitemOpen
  \bibfield  {author} {\bibinfo {author} {\bibfnamefont {Y.}~\bibnamefont {Qi}}, \bibinfo {author} {\bibfnamefont {L.~G.}\ \bibnamefont {Hector}}, \bibinfo {author} {\bibfnamefont {N.}~\bibnamefont {Ooi}}, \ and\ \bibinfo {author} {\bibfnamefont {J.~B.}\ \bibnamefont {Adams}},\ }\bibfield  {title} {\enquote {\bibinfo {title} {A first principles study of adhesion and adhesive transfer at al(111)/graphite(0001)},}\ }\href {\doibase https://doi.org/10.1016/j.susc.2005.02.048} {\bibfield  {journal} {\bibinfo  {journal} {Surface Science}\ }\textbf {\bibinfo {volume} {581}},\ \bibinfo {pages} {155--168} (\bibinfo {year} {2005})}\BibitemShut {NoStop}%
\bibitem [{\citenamefont {Lee}\ \emph {et~al.}(2008)\citenamefont {Lee}, \citenamefont {Jang}, \citenamefont {Kim},\ and\ \citenamefont {Myoung}}]{Woong}%
  \BibitemOpen
  \bibfield  {author} {\bibinfo {author} {\bibfnamefont {W.}~\bibnamefont {Lee}}, \bibinfo {author} {\bibfnamefont {S.}~\bibnamefont {Jang}}, \bibinfo {author} {\bibfnamefont {M.~J.}\ \bibnamefont {Kim}}, \ and\ \bibinfo {author} {\bibfnamefont {J.~M.}\ \bibnamefont {Myoung}},\ }\bibfield  {title} {\enquote {\bibinfo {title} {Interfacial interactions and dispersion relations in carbon–aluminium nanocomposite systems},}\ }\href {\doibase https://doi.org/10.1088/0957-4484/19/28/285701} {\bibfield  {journal} {\bibinfo  {journal} {Nanotechnology}\ }\textbf {\bibinfo {volume} {19}},\ \bibinfo {pages} {285701} (\bibinfo {year} {2008})}\BibitemShut {NoStop}%
\bibitem [{\citenamefont {Qi}\ and\ \citenamefont {Hector}(2004)}]{PhysRevB.69.235401}%
  \BibitemOpen
  \bibfield  {author} {\bibinfo {author} {\bibfnamefont {Y.}~\bibnamefont {Qi}}\ and\ \bibinfo {author} {\bibfnamefont {L.~G.}\ \bibnamefont {Hector}},\ }\bibfield  {title} {\enquote {\bibinfo {title} {Adhesion and adhesive transfer at aluminum/diamond interfaces: A first-principles study},}\ }\href {\doibase 10.1103/PhysRevB.69.235401} {\bibfield  {journal} {\bibinfo  {journal} {Phys. Rev. B}\ }\textbf {\bibinfo {volume} {69}},\ \bibinfo {pages} {235401} (\bibinfo {year} {2004})}\BibitemShut {NoStop}%
\bibitem [{\citenamefont {Tang}, \citenamefont {Sanville},\ and\ \citenamefont {Henkelman}(2009)}]{Bader}%
  \BibitemOpen
  \bibfield  {author} {\bibinfo {author} {\bibfnamefont {W.}~\bibnamefont {Tang}}, \bibinfo {author} {\bibfnamefont {E.}~\bibnamefont {Sanville}}, \ and\ \bibinfo {author} {\bibfnamefont {G.}~\bibnamefont {Henkelman}},\ }\bibfield  {title} {\enquote {\bibinfo {title} {A grid-based bader analysis algorithm without lattice bias},}\ }\href {\doibase 10.1088/0953-8984/21/8/084204} {\bibfield  {journal} {\bibinfo  {journal} {Journal of Physics: Condensed Matter}\ }\textbf {\bibinfo {volume} {21}},\ \bibinfo {pages} {084204} (\bibinfo {year} {2009})}\BibitemShut {NoStop}%
\bibitem [{\citenamefont {{K. Hindley}}(1970{\natexlab{a}})}]{RPA1}%
  \BibitemOpen
  \bibfield  {author} {\bibinfo {author} {\bibfnamefont {N.}~\bibnamefont {{K. Hindley}}},\ }\bibfield  {title} {\enquote {\bibinfo {title} {Random phase model of amorphous semiconductors i. transport and optical properties},}\ }\href {\doibase https://doi.org/10.1016/0022-3093(70)90193-6} {\bibfield  {journal} {\bibinfo  {journal} {Journal of Non-Crystalline Solids}\ }\textbf {\bibinfo {volume} {5}},\ \bibinfo {pages} {17--30} (\bibinfo {year} {1970}{\natexlab{a}})}\BibitemShut {NoStop}%
\bibitem [{\citenamefont {{K. Hindley}}(1970{\natexlab{b}})}]{RPA2}%
  \BibitemOpen
  \bibfield  {author} {\bibinfo {author} {\bibfnamefont {N.}~\bibnamefont {{K. Hindley}}},\ }\bibfield  {title} {\enquote {\bibinfo {title} {Random phase model of amorphous semiconductors ii. hot electrons},}\ }\href {\doibase https://doi.org/10.1016/0022-3093(70)90194-8} {\bibfield  {journal} {\bibinfo  {journal} {Journal of Non-Crystalline Solids}\ }\textbf {\bibinfo {volume} {5}},\ \bibinfo {pages} {31--40} (\bibinfo {year} {1970}{\natexlab{b}})}\BibitemShut {NoStop}%
\bibitem [{\citenamefont {Friedman}\ and\ \citenamefont {Mott}()}]{RPA3}%
  \BibitemOpen
  \bibfield  {author} {\bibinfo {author} {\bibfnamefont {L.}~\bibnamefont {Friedman}}\ and\ \bibinfo {author} {\bibfnamefont {N.~F.}\ \bibnamefont {Mott}},\ }\enquote {\bibinfo {title} {The hall effect near the metal–insulator transition},}\ in\ \href {\doibase 10.1142/9789812794086_0031} {\emph {\bibinfo {booktitle} {Sir Nevill Mott – 65 Years in Physics}}},\ pp.\ \bibinfo {pages} {529--534}\BibitemShut {NoStop}%
\bibitem [{\citenamefont {Friedman}(1971)}]{RPA4}%
  \BibitemOpen
  \bibfield  {author} {\bibinfo {author} {\bibfnamefont {L.}~\bibnamefont {Friedman}},\ }\bibfield  {title} {\enquote {\bibinfo {title} {Hall conductivity of amorphous semiconductors in the random phase model},}\ }\href {\doibase https://doi.org/10.1016/0022-3093(71)90024-X} {\bibfield  {journal} {\bibinfo  {journal} {Journal of Non-Crystalline Solids}\ }\textbf {\bibinfo {volume} {6}},\ \bibinfo {pages} {329--341} (\bibinfo {year} {1971})}\BibitemShut {NoStop}%
\bibitem [{\citenamefont {Mott}\ and\ \citenamefont {Davis}(1979)}]{Mott_davis}%
  \BibitemOpen
  \bibfield  {author} {\bibinfo {author} {\bibfnamefont {N.~F.}\ \bibnamefont {Mott}}\ and\ \bibinfo {author} {\bibfnamefont {E.~A.}\ \bibnamefont {Davis}},\ }\bibfield  {title} {\enquote {\bibinfo {title} {Electronic processes in non-crystalline materials},}\ \ }(\bibinfo  {publisher} {Clarendon/Oxford University Press, Oxford, New York},\ \bibinfo {year} {1979})\ \bibinfo {edition} {2nd}\ ed.,\ Chap.~\bibinfo {chapter} {2}, pp.\ \bibinfo {pages} {6--58}\BibitemShut {NoStop}%
\bibitem [{\citenamefont {Kubo}(1957)}]{Kubo}%
  \BibitemOpen
  \bibfield  {author} {\bibinfo {author} {\bibfnamefont {R.}~\bibnamefont {Kubo}},\ }\bibfield  {title} {\enquote {\bibinfo {title} {Statistical-mechanical theory of irreversible processes. i. general theory and simple applications to magnetic and conduction problems},}\ }\href {https://doi.org/10.1143/JPSJ.12.570} {\bibfield  {journal} {\bibinfo  {journal} {J.Phys.Soc.Jpn}\ }\textbf {\bibinfo {volume} {12}},\ \bibinfo {pages} {570--586} (\bibinfo {year} {1957})}\BibitemShut {NoStop}%
\bibitem [{\citenamefont {Greenwood}(1958)}]{Greenwood}%
  \BibitemOpen
  \bibfield  {author} {\bibinfo {author} {\bibfnamefont {D.~A.}\ \bibnamefont {Greenwood}},\ }\bibfield  {title} {\enquote {\bibinfo {title} {The boltzmann equation in the theory of electrical conduction in metals},}\ }\href {\doibase 10.1088/0370-1328/71/4/306} {\bibfield  {journal} {\bibinfo  {journal} {Proc.Phys.Soc.}\ }\textbf {\bibinfo {volume} {71}},\ \bibinfo {pages} {585--596} (\bibinfo {year} {1958})}\BibitemShut {NoStop}%
\bibitem [{\citenamefont {Moseley}\ and\ \citenamefont {Lukes}(1978)}]{Mosely&Lukes}%
  \BibitemOpen
  \bibfield  {author} {\bibinfo {author} {\bibfnamefont {L.~L.}\ \bibnamefont {Moseley}}\ and\ \bibinfo {author} {\bibfnamefont {T.}~\bibnamefont {Lukes}},\ }\bibfield  {title} {\enquote {\bibinfo {title} {{A simplified derivation of the Kubo‐Greenwood formula}},}\ }\href {\doibase 10.1119/1.11229} {\bibfield  {journal} {\bibinfo  {journal} {American Journal of Physics}\ }\textbf {\bibinfo {volume} {46}},\ \bibinfo {pages} {676--677} (\bibinfo {year} {1978})}\BibitemShut {NoStop}%
\bibitem [{\citenamefont {Subedi}\ \emph {et~al.}(2022)\citenamefont {Subedi}, \citenamefont {Kappagantula}, \citenamefont {Kraft}, \citenamefont {Nittala},\ and\ \citenamefont {Drabold}}]{K1}%
  \BibitemOpen
  \bibfield  {author} {\bibinfo {author} {\bibfnamefont {K.~N.}\ \bibnamefont {Subedi}}, \bibinfo {author} {\bibfnamefont {K.}~\bibnamefont {Kappagantula}}, \bibinfo {author} {\bibfnamefont {F.}~\bibnamefont {Kraft}}, \bibinfo {author} {\bibfnamefont {A.}~\bibnamefont {Nittala}}, \ and\ \bibinfo {author} {\bibfnamefont {D.~A.}\ \bibnamefont {Drabold}},\ }\bibfield  {title} {\enquote {\bibinfo {title} {Electrical conduction processes in aluminum: Defects and phonons},}\ }\href {\doibase 10.1103/PhysRevB.105.104114} {\bibfield  {journal} {\bibinfo  {journal} {Phys. Rev. B}\ }\textbf {\bibinfo {volume} {105}},\ \bibinfo {pages} {104114} (\bibinfo {year} {2022})}\BibitemShut {NoStop}%
\bibitem [{\citenamefont {Subedi}\ \emph {et~al.}(2019)\citenamefont {Subedi}, \citenamefont {Prasai}, \citenamefont {Kozicki},\ and\ \citenamefont {Drabold}}]{K2}%
  \BibitemOpen
  \bibfield  {author} {\bibinfo {author} {\bibfnamefont {K.~N.}\ \bibnamefont {Subedi}}, \bibinfo {author} {\bibfnamefont {K.}~\bibnamefont {Prasai}}, \bibinfo {author} {\bibfnamefont {M.~N.}\ \bibnamefont {Kozicki}}, \ and\ \bibinfo {author} {\bibfnamefont {D.~A.}\ \bibnamefont {Drabold}},\ }\bibfield  {title} {\enquote {\bibinfo {title} {Structural origins of electronic conduction in amorphous copper-doped alumina},}\ }\href {\doibase 10.1103/PhysRevMaterials.3.065605} {\bibfield  {journal} {\bibinfo  {journal} {Phys. Rev. Materials}\ }\textbf {\bibinfo {volume} {3}},\ \bibinfo {pages} {065605} (\bibinfo {year} {2019})}\BibitemShut {NoStop}%
\bibitem [{\citenamefont {Subedi}, \citenamefont {Prasai},\ and\ \citenamefont {Drabold}(2020)}]{K4}%
  \BibitemOpen
  \bibfield  {author} {\bibinfo {author} {\bibfnamefont {K.~N.}\ \bibnamefont {Subedi}}, \bibinfo {author} {\bibfnamefont {K.}~\bibnamefont {Prasai}}, \ and\ \bibinfo {author} {\bibfnamefont {D.~A.}\ \bibnamefont {Drabold}},\ }\bibfield  {title} {\enquote {\bibinfo {title} {Space-projected conductivity and spectral properties of the conduction matrix},}\ }\href {\doibase https://doi.org/10.1002/pssb.202000438} {\bibfield  {journal} {\bibinfo  {journal} {Phys. Status Solidi B}\ }\textbf {\bibinfo {volume} {258}},\ \bibinfo {pages} {2000438} (\bibinfo {year} {2020})}\BibitemShut {NoStop}%
\bibitem [{\citenamefont {Ugwumadu}\ \emph {et~al.}(2023{\natexlab{a}})\citenamefont {Ugwumadu}, \citenamefont {Nepal}, \citenamefont {Thapa}, \citenamefont {Lee}, \citenamefont {{Al Majali}}, \citenamefont {Trembly},\ and\ \citenamefont {Drabold}}]{aF}%
  \BibitemOpen
  \bibfield  {author} {\bibinfo {author} {\bibfnamefont {C.}~\bibnamefont {Ugwumadu}}, \bibinfo {author} {\bibfnamefont {K.}~\bibnamefont {Nepal}}, \bibinfo {author} {\bibfnamefont {R.}~\bibnamefont {Thapa}}, \bibinfo {author} {\bibfnamefont {Y.}~\bibnamefont {Lee}}, \bibinfo {author} {\bibfnamefont {Y.}~\bibnamefont {{Al Majali}}}, \bibinfo {author} {\bibfnamefont {J.}~\bibnamefont {Trembly}}, \ and\ \bibinfo {author} {\bibfnamefont {D.}~\bibnamefont {Drabold}},\ }\bibfield  {title} {\enquote {\bibinfo {title} {Simulation of multi-shell fullerenes using machine-learning gaussian approximation potential},}\ }\href {\doibase https://doi.org/10.1016/j.cartre.2022.100239} {\bibfield  {journal} {\bibinfo  {journal} {Carbon Trends}\ }\textbf {\bibinfo {volume} {10}},\ \bibinfo {pages} {100239} (\bibinfo {year} {2023}{\natexlab{a}})}\BibitemShut {NoStop}%
\bibitem [{\citenamefont {Ugwumadu}\ \emph {et~al.}(2023{\natexlab{b}})\citenamefont {Ugwumadu}, \citenamefont {Thapa}, \citenamefont {Al-Majali}, \citenamefont {Trembly},\ and\ \citenamefont {Drabold}}]{nt}%
  \BibitemOpen
  \bibfield  {author} {\bibinfo {author} {\bibfnamefont {C.}~\bibnamefont {Ugwumadu}}, \bibinfo {author} {\bibfnamefont {R.}~\bibnamefont {Thapa}}, \bibinfo {author} {\bibfnamefont {Y.}~\bibnamefont {Al-Majali}}, \bibinfo {author} {\bibfnamefont {J.}~\bibnamefont {Trembly}}, \ and\ \bibinfo {author} {\bibfnamefont {D.~A.}\ \bibnamefont {Drabold}},\ }\bibfield  {title} {\enquote {\bibinfo {title} {Formation of amorphous carbon multi-walled nanotubes from random initial configurations},}\ }\href {\doibase https://doi.org/10.1002/pssb.202200527} {\bibfield  {journal} {\bibinfo  {journal} {physica status solidi (b)}\ }\textbf {\bibinfo {volume} {260}},\ \bibinfo {pages} {2200527} (\bibinfo {year} {2023}{\natexlab{b}})}\BibitemShut {NoStop}%
\bibitem [{\citenamefont {Ugwumadu}\ \emph {et~al.}(2023{\natexlab{c}})\citenamefont {Ugwumadu}, \citenamefont {Thapa}, \citenamefont {Nepal},\ and\ \citenamefont {Drabold}}]{aG2}%
  \BibitemOpen
  \bibfield  {author} {\bibinfo {author} {\bibfnamefont {C.}~\bibnamefont {Ugwumadu}}, \bibinfo {author} {\bibfnamefont {R.}~\bibnamefont {Thapa}}, \bibinfo {author} {\bibfnamefont {K.}~\bibnamefont {Nepal}}, \ and\ \bibinfo {author} {\bibfnamefont {D.~A.}\ \bibnamefont {Drabold}},\ }\bibfield  {title} {\enquote {\bibinfo {title} {Atomistic nature of amorphous graphite},}\ }\href {\doibase 10.13036/17533562.64.1.18} {\bibfield  {journal} {\bibinfo  {journal} {European Journal of Glass Science and Technology Part B}\ }\textbf {\bibinfo {volume} {64}},\ \bibinfo {pages} {16--22} (\bibinfo {year} {2023}{\natexlab{c}})}\BibitemShut {NoStop}%
\bibitem [{\citenamefont {Thapa}\ \emph {et~al.}(2022)\citenamefont {Thapa}, \citenamefont {Ugwumadu}, \citenamefont {Nepal}, \citenamefont {Trembly},\ and\ \citenamefont {Drabold}}]{prl_raj}%
  \BibitemOpen
  \bibfield  {author} {\bibinfo {author} {\bibfnamefont {R.}~\bibnamefont {Thapa}}, \bibinfo {author} {\bibfnamefont {C.}~\bibnamefont {Ugwumadu}}, \bibinfo {author} {\bibfnamefont {K.}~\bibnamefont {Nepal}}, \bibinfo {author} {\bibfnamefont {J.}~\bibnamefont {Trembly}}, \ and\ \bibinfo {author} {\bibfnamefont {D.~A.}\ \bibnamefont {Drabold}},\ }\bibfield  {title} {\enquote {\bibinfo {title} {Ab initio simulation of amorphous graphite},}\ }\href {\doibase 10.1103/PhysRevLett.128.236402} {\bibfield  {journal} {\bibinfo  {journal} {Phys. Rev. Lett.}\ }\textbf {\bibinfo {volume} {128}},\ \bibinfo {pages} {236402} (\bibinfo {year} {2022})}\BibitemShut {NoStop}%
\bibitem [{\citenamefont {Ono}(1976)}]{ONO}%
  \BibitemOpen
  \bibfield  {author} {\bibinfo {author} {\bibfnamefont {S.}~\bibnamefont {Ono}},\ }\bibfield  {title} {\enquote {\bibinfo {title} {C-axis resistivity of graphite in connection with stacking faults},}\ }\href {\doibase 10.1143/JPSJ.40.498} {\bibfield  {journal} {\bibinfo  {journal} {Journal of the Physical Society of Japan}\ }\textbf {\bibinfo {volume} {40}},\ \bibinfo {pages} {498--504} (\bibinfo {year} {1976})}\BibitemShut {NoStop}%
\bibitem [{\citenamefont {Iwashita}, \citenamefont {Imagawa},\ and\ \citenamefont {Nishiumi}(2013)}]{Norio_graphite}%
  \BibitemOpen
  \bibfield  {author} {\bibinfo {author} {\bibfnamefont {N.}~\bibnamefont {Iwashita}}, \bibinfo {author} {\bibfnamefont {H.}~\bibnamefont {Imagawa}}, \ and\ \bibinfo {author} {\bibfnamefont {W.}~\bibnamefont {Nishiumi}},\ }\bibfield  {title} {\enquote {\bibinfo {title} {Variation of temperature dependence of electrical resistivity with crystal structure of artificial graphite products},}\ }\href {\doibase 10.1016/J.CARBON.2013.05.042} {\bibfield  {journal} {\bibinfo  {journal} {carbon}\ }\textbf {\bibinfo {volume} {61}},\ \bibinfo {pages} {602--608} (\bibinfo {year} {2013})}\BibitemShut {NoStop}%
\bibitem [{\citenamefont {Bapat}(1973)}]{Bapat_carbon}%
  \BibitemOpen
  \bibfield  {author} {\bibinfo {author} {\bibfnamefont {S.}~\bibnamefont {Bapat}},\ }\bibfield  {title} {\enquote {\bibinfo {title} {Thermal conductivity and electrical resistivity of two types of {ATJ-S} graphite to 3500° k},}\ }\href {\doibase https://doi.org/10.1016/0008-6223(73)90310-2} {\bibfield  {journal} {\bibinfo  {journal} {carbon}\ }\textbf {\bibinfo {volume} {11}},\ \bibinfo {pages} {511--514} (\bibinfo {year} {1973})}\BibitemShut {NoStop}%
\bibitem [{\citenamefont {Matsubara}, \citenamefont {Sugihara},\ and\ \citenamefont {Tsuzuku}(1990)}]{matsubara}%
  \BibitemOpen
  \bibfield  {author} {\bibinfo {author} {\bibfnamefont {K.}~\bibnamefont {Matsubara}}, \bibinfo {author} {\bibfnamefont {K.}~\bibnamefont {Sugihara}}, \ and\ \bibinfo {author} {\bibfnamefont {T.}~\bibnamefont {Tsuzuku}},\ }\bibfield  {title} {\enquote {\bibinfo {title} {Electrical resistance in the c direction of graphite},}\ }\href {\doibase 10.1103/PhysRevB.41.969} {\bibfield  {journal} {\bibinfo  {journal} {Phys. Rev. B}\ }\textbf {\bibinfo {volume} {41}},\ \bibinfo {pages} {969--974} (\bibinfo {year} {1990})}\BibitemShut {NoStop}%
\end{thebibliography}%

\end{document}


\preprint{APS/123-QED}

\title{Supplementary Material \\ Physical origin of enhanced electrical conduction in aluminum-graphene composites}%

\author{K. Nepal}
\email{kn478619@ohio.edu}
\affiliation{Department of Physics and Astronomy, Nanoscale and Quantum Phenomena Institute (NQPI)
Ohio University, Athens, Ohio 45701, USA}%

\author{C. Ugwumadu}
\affiliation{Department of Physics and Astronomy, Nanoscale and Quantum Phenomena Institute (NQPI)
Ohio University, Athens, Ohio 45701, USA}%

\author{K.N. Subedi}
\affiliation{Theoretical Division, Los Alamos National Laboratory, Los Alamos, New Mexico 87545, USA}%

\author{K. Kappagantula}%
\affiliation{Pacific Northwest National Laboratory, 
Richland, Washington, 99352, USA}%

\author{D. A. Drabold}%
\email{drabold@ohio.edu}
\affiliation{Department of Physics and Astronomy, Nanoscale and Quantum Phenomena Institute (NQPI)
Ohio University, Athens, Ohio 45701, USA}%

\date{\today}

\begin{abstract}
The electronic and transport properties of aluminum-graphene composite materials were investigated using \textit{ab initio} plane wave density functional theory.  The interfacial structure is reported for several configurations. In some cases, the face-centered aluminum (111) surface relaxes in a nearly ideal registry with graphene, resulting in a remarkably continuous interface structure. The Kubo-Greenwood formula and space-projected conductivity were employed to study electronic conduction in aluminum single- and double-layer graphene-aluminum composite models. The electronic density of states at the Fermi level is enhanced by the graphene for certain aluminum-graphene interfaces, thus, improving electronic conductivity. In double-layer graphene composites, conductivity varies non-monotonically with temperature, showing an increase between 300-400 K at short aluminum-graphene distances, unlike the consistent decrease in single-layer composites.
\end{abstract}


\maketitle

\section{Simulation Protocol Utilized in VASP for Calculations in this Work}
All the calculations in this work were performed using the plane wave density functional theory code,  VASP (Vienna \textit{ab initio} simulation package) \cite{VASP}. For geometry optimization using conjugate gradient in VASP (maximum residual force is less than 0.01 eV/$\AA$), we used a kinetic energy cutoff of 420 eV. For single-point calculations, we used an energy cutoff of 480 eV. Projected augmented wave (PAW) potentials were implemented for ion–election interactions, and the generalized gradient approximation (GGA) of Perdew–Burke–Ernzerhof (PBE) as the exchange-correlation functional \cite{PAW, PBE}. The Brillouin zone was sampled using the Monkhorst-Pack \cite{monkhorst1976special} scheme with 2 $\times$ 2 $\times$ 1 k-point meshes as implemented in VASP.

\section{Pressure-relaxed Al-G Composite Models}
To impose compression onto the system, the vertical dimension of the composite model was reduced followed by a structural relaxation. For each compression followed by relaxation, the initial and final interfacial distances were noted and are listed in Table \ref{tab:KModels}, along with the external pressure on relaxed models computed within VASP.  Table \ref{tab:KModels} left and right correspond to SL and DL composites. The graphene layer formed on either side was at similar configurations with the interfacial Al layers after relaxation. 

\begin{table*}[h!]
         \caption{Summary of variation in the interfacial distance for different Al-G models after atomic structure relaxation and corresponding external pressure. The left and right correspond to SL and DL graphene aluminum composites respectively.}
        \label{tab:KModels}
		\begin{tabular*}{\linewidth}{@{\extracolsep\fill}cccc|cccc} 
            
              &  \textbf{Single}  &   &   &  & \textbf{Double}  &    \\
                \hline
             \rowcolor{lightgray!20}  \footnote{\label{note2} Reduced Al-G interfacial distance before conjugate gradient relaxation in VASP} $d_{Al-G}$ &  \footnote{\label{note3} Obtained Al-G interfacial distance after conjugate gradient relaxation was completed} $d_{Al-G}$  & \footnote{\label{note1} External pressure computed for the compressed models within VASP.} P & & \textsuperscript{\ref{note2}} $d_{Al-G}$ & \textsuperscript{\ref{note3}} $d_{Al-G}$  &  \textsuperscript{\ref{note1}} P \\
             
             [$\AA$] & [$\AA$] &  [kB] & & [$\AA$]& [$\AA$]  & [kB]  \\
             \hline
             
 		 3.48 & 3.40 &  0.27 & & 3.48 & 3.41  & 1.15   \\ 
              3.42 & 3.35  &  3.42 & & 3.42  & 3.35   & 2.77     \\
		   3.36 & 3.31 & 8.41 & & 3.36 & 3.31  & 6.25     \\
		   3.31  & 3.25  & 16.82 & & 3.30 & 3.24  & 9.72  \\
              3.28 & 3.13  &  20.09 & & 3.24 & 3.19  & 14.35 \\
		   3.21 & 3.01  & 27.29 & &  3.18 & 3.11 & 17.50 \\
              3.16 & 2.90  & 36.68 & & 3.12  & 3.07  & 22.66   \\
		   3.09 & 2.71  & 69.76 & & 3.06  & 3.01  & 27.64  \\
                   &       &       & & 3.00  &  2.97 & 31.88  \\
		         &        &       & & 2.96  &  2.94  & 34.69   \\
		\end{tabular*}
\end{table*}

\section{{Simulation protocol for Temperature-dependent Conductivity Calculations}}\label{sec:Tdependent}

To perform the temperature-dependent conductivity calculations, we followed the simulation protocol outlined in reference \cite{K1}. Selected models were equilibrated in a canonical ensemble for a temperature range between 100 K to 600 K, in steps of 50 K. The temperature was maintained using the Nos\'e-Hoover thermostat. The simulations were conducted for 3 ps with a timestep of 1.5 fs. For the subsequent conductivity calculations, we selected the last 10 configurations separated by intervals of 50 fs for each temperature considered. At temperatures below the Debye temperature, classical MD is not justified, as lattice quantization should be considered; however, while the dynamics are unrealistic it appears that the naive classical sampling yields sensible results when employed with KGF, as we demonstrated and discussed in detail earlier and discussed in detail \cite{K1}.

 \begin{figure}[h!]
        \centering
	\includegraphics[width=.7\linewidth]{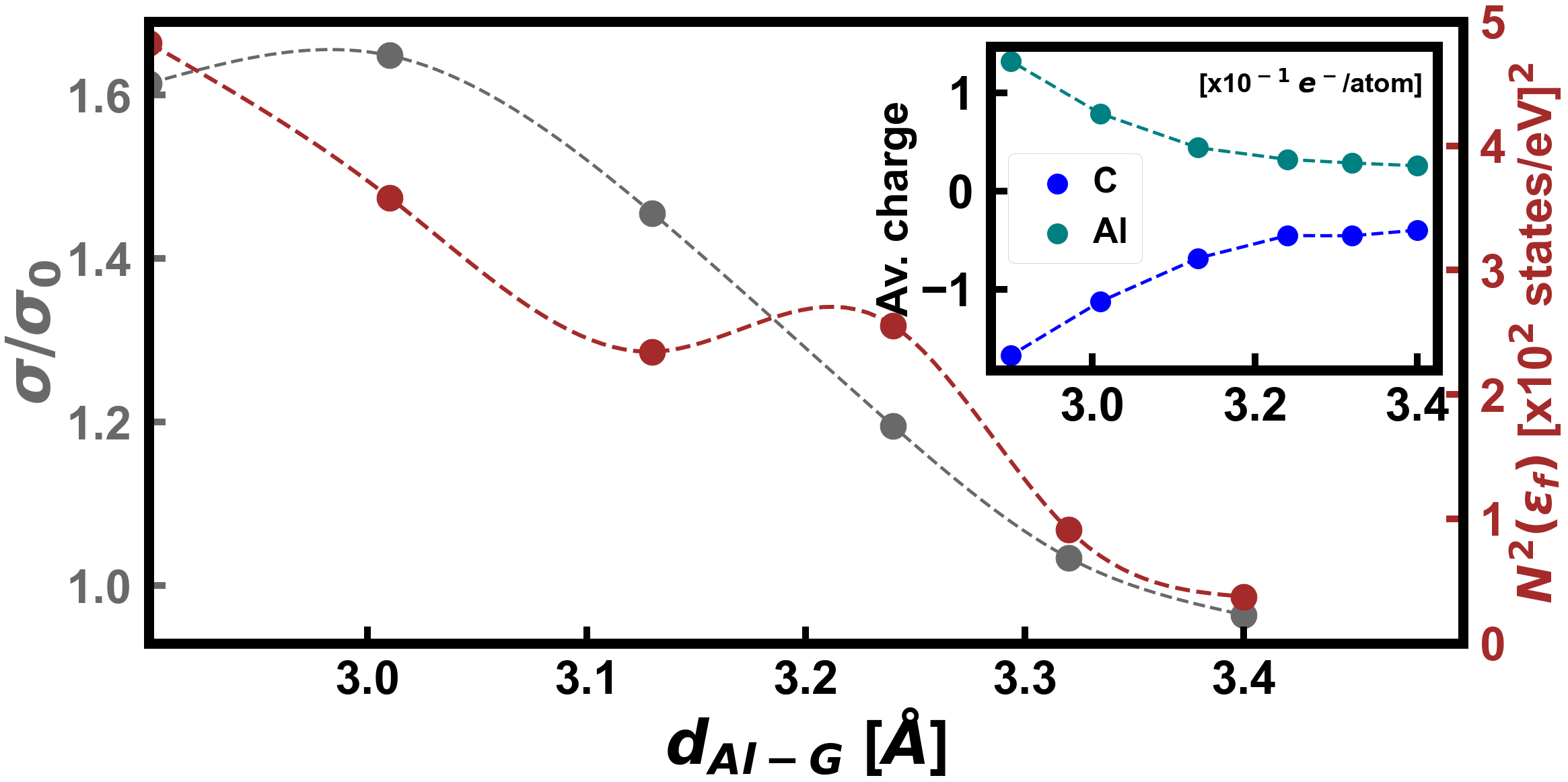}
         \caption{ All plot corresponds to the SL model for the different Al-G distances ($d_{Al-G}$) in the x-axis. The conductivity of the relaxed SL models is represented by the gray curve, with $\sigma_{0}$ denoting the conductivity of the Al-matrix calculated at 300K. The squared density of states at the Fermi level is shown in brown. In the inset, Bader analysis illustrates the average charge gain and loss for C (blue) and Al (green) atoms. Dotted lines are included in all plots as visual aids. }

	\label{fig:KSfiig_KFGBBader}
\end{figure}

\newpage
\bibliography{APL_AL_G_biblography}